\documentclass[aps,prd,a4paper,preprintnumbers,tightenlines,superscriptaddress,
twocolumn,showpacs,final,nofootinbib,floatfix]{revtex4-1}
\pdfoutput=1
\usepackage{graphicx}
\usepackage{longtable}
\usepackage{amsmath}
\usepackage{amssymb}
\usepackage{hyperref}
\usepackage{dcolumn}
\usepackage{bm}
\usepackage{multirow}
\usepackage{cleveref}
\usepackage{bm}
\usepackage{ wasysym }
\usepackage{color}
\usepackage{stackengine}
\usepackage[export]{adjustbox}

\newcommand{\be}{\begin{eqnarray}}
\newcommand{\ee}{\end{eqnarray}}
\newcommand{\mpl}{M_{\rm {pl}}}

\newcommand{\dd}{\, {\rm d}}
\newcommand{\gsim}{\;\mbox{\raisebox{-0.5ex}{$\stackrel{>}{\scriptstyle{\sim}}$}
}\;}
\newcommand{\lsim}{\;\mbox{\raisebox{-0.5ex}{$\stackrel{<}{\scriptstyle{\sim}}$}
}\;}
\def\eea{\end{eqnarray}}
\def\bea{\begin{eqnarray}}

\newcommand{\rs}{r_{\rm s}}
\newcommand{\pn}{\Phi_{\rm N}}

\newcommand{\GP}{G^{\rm PPN}}

\newcommand{\tg}{\tilde{g}}

\newcommand{\eff}{_{\rm eff}}

\newcommand{\nm}{{\mu\nu}}

\newcommand{\pmi}{\phi_{\rm min}}

\newcommand{\GN}{G_{\rm N}}
\newcommand{\mmm}{{_{\rm m}}}

\newcommand{\oo}{\mathcal{O}}

\newcommand{\rv}{r_{\rm V}}

\usepackage{titlesec}
\titleclass{\subsubsubsection}{straight}[\subsection]

\titleformat{\paragraph}
{\normalfont\normalsize\bfseries}{\theparagraph}{1em}{}
\titlespacing*{\paragraph}{0pt}{3.25ex plus 1ex minus .2ex}{1.5ex plus .2ex}

\newcounter{subsubsubsection}[subsubsection]
\renewcommand\thesubsubsubsection{\thesubsubsection.\arabic{subsubsubsection}}
\renewcommand\theparagraph{\thesubsubsubsection.\arabic{paragraph}} % optional; useful if paragraphs are to be numbered

\titleformat{\subsubsubsection}
  {\normalfont\normalsize\bfseries}{\thesubsubsubsection}{1em}{}
\titlespacing*{\subsubsubsection}
{0pt}{3.25ex plus 1ex minus .2ex}{1.5ex plus .2ex}

\makeatletter
\renewcommand\paragraph{\@startsection{paragraph}{5}{\z@}%
  {3.25ex \@plus1ex \@minus.2ex}%
  {-1em}%
  {\normalfont\normalsize\bfseries}}
\renewcommand\subparagraph{\@startsection{subparagraph}{6}{\parindent}%
  {3.25ex \@plus1ex \@minus .2ex}%
  {-1em}%
  {\normalfont\normalsize\bfseries}}
\def\toclevel@subsubsubsection{4}
\def\toclevel@paragraph{5}
\def\toclevel@paragraph{6}
\def\l@subsubsubsection{\@dottedtocline{4}{7em}{4em}}
\def\l@paragraph{\@dottedtocline{5}{10em}{5em}}
\def\l@subparagraph{\@dottedtocline{6}{14em}{6em}}
\makeatother

\begin{document}
\title{Tests of Gravity with Future Space-Based Experiments}
\author{Jeremy Sakstein}
\affiliation{Center for Particle Cosmology, Department of Physics and Astronomy, University of Pennsylvania 209 S. 33rd St., Philadelphia, PA 19104, USA}
\email{sakstein@physics.upenn.edu}

\begin{abstract}
Future space-based tests of relativistic gravitation---laser ranging to Phobos, accelerometers in orbit, and optical networks surrounding Earth---will constrain the theory of gravity with unprecedented precision by testing the inverse-square law, the strong and weak equivalence principles, and the deflection and time delay of light by massive bodies. In this paper, we estimate the bounds that could be obtained on alternative gravity theories that use screening mechanisms to suppress deviations from general relativity in the solar system: chameleon, symmetron, and galileon models. We find that space-based tests of the parameterized post-Newtonian parameter $\gamma$ will constrain chameleon and symmetron theories to new levels, and that tests of the inverse-square law using laser ranging to Phobos will provide the most stringent constraints on galileon theories to date. We end by discussing the potential for constraining these theories using upcoming tests of the weak equivalence principle, and conclude that further theoretical modeling is required in order to fully utilize the data.
\end{abstract}
\maketitle

\section{Introduction}

We are in a golden age for testing relativistic theories of gravitation. The recent discovery of gravitational waves from merging binary black holes \cite{Abbott:2016blz} has tested general relativity (GR) in the strong field regime for the first time\footnote{See \cite{Sakstein:2017xjx,Creminelli:2017sry,Ezquiaga:2017ekz,Baker:2017hug,Crisostomi:2017lbg,Langlois:2017dyl,Kreisch:2017uet} for tests of cosmological infra-red modifications of gravity using the recent simultaneous observation of both gravitational waves (GW170817) and a gamma ray burst (GRB 170817A) from merging neutron stars by the LIGO/Virgo and Fermi collaborations. } \cite{TheLIGOScientific:2016src}, and, on cosmological scales, ongoing surveys such as DES, as well as future surveys such as Euclid, LSST, SKA, and WFIRST will test the theory of gravity on cosmological distance scales \cite{Amendola:2012ys}. On Earth, advances in table-top experiments such as torsion-balance experiments \cite{Kapner:2006si}, optically levitated microspheres \cite{Rider:2016xaq}, and atom interferometry \cite{Jaffe:2016fsh} have probed new potential gravitational interactions at micron distances, and forces as weak as $10^{-18}N$. 

From a theoretical viewpoint, there has been a resurgence in the study of modified gravity models driven by the mysterious acceleration of the cosmic expansion: dark energy \cite{Copeland:2006wr,Clifton:2011jh,Joyce:2014kja,Koyama:2015vza}. Typically, cosmologically relevant modifications of gravity are difficult to reconcile with solar system tests of GR, either because they require strong couplings to matter or because they have force ranges of order the size of the universe. This has led the community to focus on a narrow class of models that include \emph{screening mechanisms} \cite{Jain:2010ka,Khoury:2010xi,Joyce:2014kja,Burrage:2017qrf}. Screening mechanisms use non-linear effects to suppress deviations from GR in the solar system while allowing them to be relevant on larger, cosmological scales. For this reason, the free parameters (masses and couplings) do not need to be tuned to evade solar system tests.

Complementary to the tests mentioned above, the next generation of space-based tests\footnote{See \cite{Turyshev:2008ur,Turyshev:2008dr,Turyshev:2007qy} for recent reviews.}---accelerometers in orbit, laser networks surrounding the Sun and Earth, and laser ranging to Mars---will constrain relativistic gravity to unprecedented levels in the solar system by measuring the parameters $\gamma$, $\beta$, and $\delta$ appearing in the parameterized post-Newtonian (PPN) metric, testing the strong and weak equivalence principles, constraining the time-variation of Newton's constant, and by looking for deviations in the inverse-square law. The purpose of this paper is to explore the implications of these future missions for three theories of gravity that exhibit different screening mechanisms: chameleon \cite{Khoury:2003aq,Khoury:2003rn}, symmetron \cite{Hinterbichler:2010es}, and galileon \cite{Nicolis:2008in} theories.

We will proceed as follows: In the next section, we will motivate screening mechanisms and introduce the three mentioned above. Next, in section \ref{sec:missions} we will briefly review the current missions that have tested gravity in space, and the proposed missions that we will use in this work to forecast the projected bounds on the model's parameter space. In section \ref{sec:constraints} we will present the current and projected bounds, and discuss their implications for the models, and for other tests of screening mechanisms. We will also discuss other future tests that may be useful for testing screening mechanisms but that we will not forecast for here due to uncertainties in the theoretical modeling; we discuss these in order to highlight how a dedicated effort towards a better modeling of these systems could improve the current bounds on screened modified gravity models. We conclude in section \ref{sec:concs}. In Appendix \ref{sec:appendix} we provide a brief derivation of the PPN parameter for chameleon and symmetron theories. 

\section{Screening Mechanisms}
\label{sec:screening_mechanisms}

\subsection{Why Screening?}

The study of scalar-tensor theories has been motivated by the cosmological observation of dark energy, the mysterious driving mechanism for the acceleration of the cosmic expansion. Indeed, one proposed explanation is that gravity is modified on large distances. In order to be relevant today, any modification must necessarily be as important as general relativity but this cannot be the case in the solar system because deviations are constrained to be subdominant by a factor of $10^{-5}$ or more depending on the specific theory\footnote{For example, a theory that predicts strong violations of the weak equivalence principle will be constrained to levels of $\oo(10^{-15})$ \cite{Touboul:2017grn,Berge:2017ovy}. }. As an example, consider Brans-Dicke gravity, which describes a new scalar field $\phi$ coupled to gravity and is parameterized by a single parameter $\omega_{\rm BD}$. In the non-relativistic limit, one finds a Poisson-like equation for $\phi$:
\begin{equation}\label{eq:poissonBD}
\nabla^2\phi=-\frac{8\pi G}{2+3\omega_{\rm BD}}\rho,
\end{equation}
which gives a contribution to the PPN parameter $\gamma$
\begin{equation}
|\gamma-1|=\frac{1}{2+\omega_{\rm BD}}.
\end{equation}
In order to satisfy the Cassini bound $|\gamma-1|<2.1\times10^{-5}$ \cite{Bertotti:2003rm} one needs to take $\omega_{\rm BD}>40000$, but, examining equation \eqref{eq:poissonBD} one can see that the effective coupling to matter $\alpha_{\rm eff}\sim 1/\omega_{\rm BD}\lsim 10^{-4}$. This implies that any Brans-Dicke-like modifications of GR must be subdominant to the Einstein-Hilbert term on all scales by at least a factor of $10^4$. Such a requirement means that any such theories are cosmologically irrelevant.

One reason that solar system tests are so constraining for Brans-Dicke-like theories is that they contain massless scalars, and hence fit into the PPN form due to the resultant $1/r$ potentials. One can try to circumvent this issue but introducing a mass for the scalar so that its equation of motion is
\begin{equation}
\label{eq:massivescalar}
(\nabla^2+m^2)\phi=8\pi\alpha G \rho,
\end{equation}
in which case the total potential sourced by a static, spherically symmetric body is of the Yukawa form
\begin{equation}
V(r)=\frac{GM}{r}(1+2\alpha^2 e^{-mr}).
\end{equation}
Yuakawa forces have been searched for extensively at distances ranging from the Earth-Moon distance \cite{Merkowitz:2010kka,Murphy:2012rea} to micron-scales \cite{Adelberger:2003zx,Adelberger:2005vu,Kapner:2006si}, and so the mass $m>(\mu\textrm{m})^{-1}$ in order to evade these tests. Again, such a scalar can have nothing to say about cosmological-scale physics.

One common issue with the previous two models is that solar system tests automatically preclude any relevance for cosmology because the force must either be too weak, or too short-ranged. Screening mechanisms circumvent this problem by introducing non-linear modifications of the Poisson equation that dynamically suppress deviations from GR in the solar system without the need to fine-tune the mass or the coupling to matter. In this paper we will consider three well-studied screening mechanisms:

\begin{itemize}
\item {\bf Chameleon Screening:} This dynamically changes the mass of the field so that it mediates a short ranged force in the solar system but may influence cosmology on Mpc scales.
\item {\bf Symmetron Screening:} This dynamically varies the coupling to matter so that it is essentially uncoupled in the solar system but can source deviations from GR on linear cosmological scales.
\item {\bf The Vainshtein Mechanism:} This uses non-linear kinetic terms to alter the field profile sourced by massive bodies so that fifth-forces are highly-suppressed in the solar system. On cosmological scales, theories that exhibit this mechanism can self-accelerate without a cosmological constant, which makes them interesting alternatives to $\Lambda$CDM. The fifth-forces can also modify the dynamics of linear and non-linear perturbations \cite{Barreira:2012kk,Barreira:2013eea,Barreira:2013xea}.
\end{itemize}

We now proceed to discuss each of these briefly in turn. Our discussion will be far from comprehensive and the interested reader is directed to references \cite{Hui:2009kc,Sakstein:2015oqa,Burrage:2016bwy,Babichev:2013usa,Burrage:2017qrf} for more details\footnote{Unpublished lecture notes can be found at the following \href{http://www.jeremysakstein.com/astro_grav_2.pdf}{url}. }.

\subsection{Screening Mechanisms}

\subsubsection{Chameleon Screening}

Chameleon screening \cite{Khoury:2003aq,Khoury:2003rn} uses a non-linear potential to make the field's mass a function of the environmental density. It's equation of motion is\footnote{{Note that we have switched to a dimensionful scalar in keeping with the conventions in the literature. This is why there is a factor of $\alpha\rho/\mpl$ rather than $8\pi\alpha G\rho$ as in equation \eqref{eq:massivescalar}, which used a dimensionless scalar to ensure that the equation had a similar form to the Poisson equation in GR.}}
\begin{equation}
\nabla^2\phi=-n\frac{\Lambda^{4+n}}{\phi^{n+1}}+\frac{\alpha\rho}{\mpl},
\end{equation}
the right hand side of which can be derived from an effective potential
\begin{equation}
V\eff=\frac{\Lambda^{4+n}}{\phi^{n}}+\frac{\alpha\phi\rho}{\mpl}.
\end{equation}
The mass-scale $\Lambda$ can vary over many orders of magnitude, but it is often compared to the dark energy scale $\Lambda_{\rm DE}=2.4$ meV since this value is relevant for the present-day cosmic acceleration\footnote{As an example, many authors consider a generalized potential of the form $V(\phi)=\Lambda^4\exp(\Lambda^n/\phi^n)=\Lambda^4+\Lambda^{4+n}/\phi^n+\cdots$, which would give a common origin for the cosmological constant and the chameleon. Note that the chameleon cannot accelerate cosmologically without a cosmological constant \cite{Wang:2012kj}.}. The location of the minimum of the effective potential is density-dependent
\begin{equation}\label{eqSM:phimincam}
\pmi(\rho)=\left(\frac{n\mpl\Lambda^{4+n}}{\alpha\rho}\right)^{\frac{1}{n+1}},
\end{equation}
and hence so is the effective mass of the field about said minimum
\begin{equation}\label{eq:eqSM:meffcham}
m_{\rm eff}^2=V''_{\rm eff}(\phi)=n(n+1)\Lambda^{n+4}\left(\frac{\alpha\rho}{n\mpl\Lambda^{n+4}}\right)^{\frac{n+2}{n+1}}.
\end{equation}
Since the cosmological and terrestrial density vary by 29 orders of magnitude, the parameters can be chosen such that the chameleon force in laboratory experiments is sub-micron. Current experimental searches \cite{Burrage:2016bwy,Burrage:2017qrf} imply that the chameleon cannot drive the cosmic acceleration \cite{Wang:2012kj} but the chameleon force can still be relevant for cosmology on smaller (Mpc) scales. 

\begin{figure}
{\includegraphics[width=0.4\textwidth]{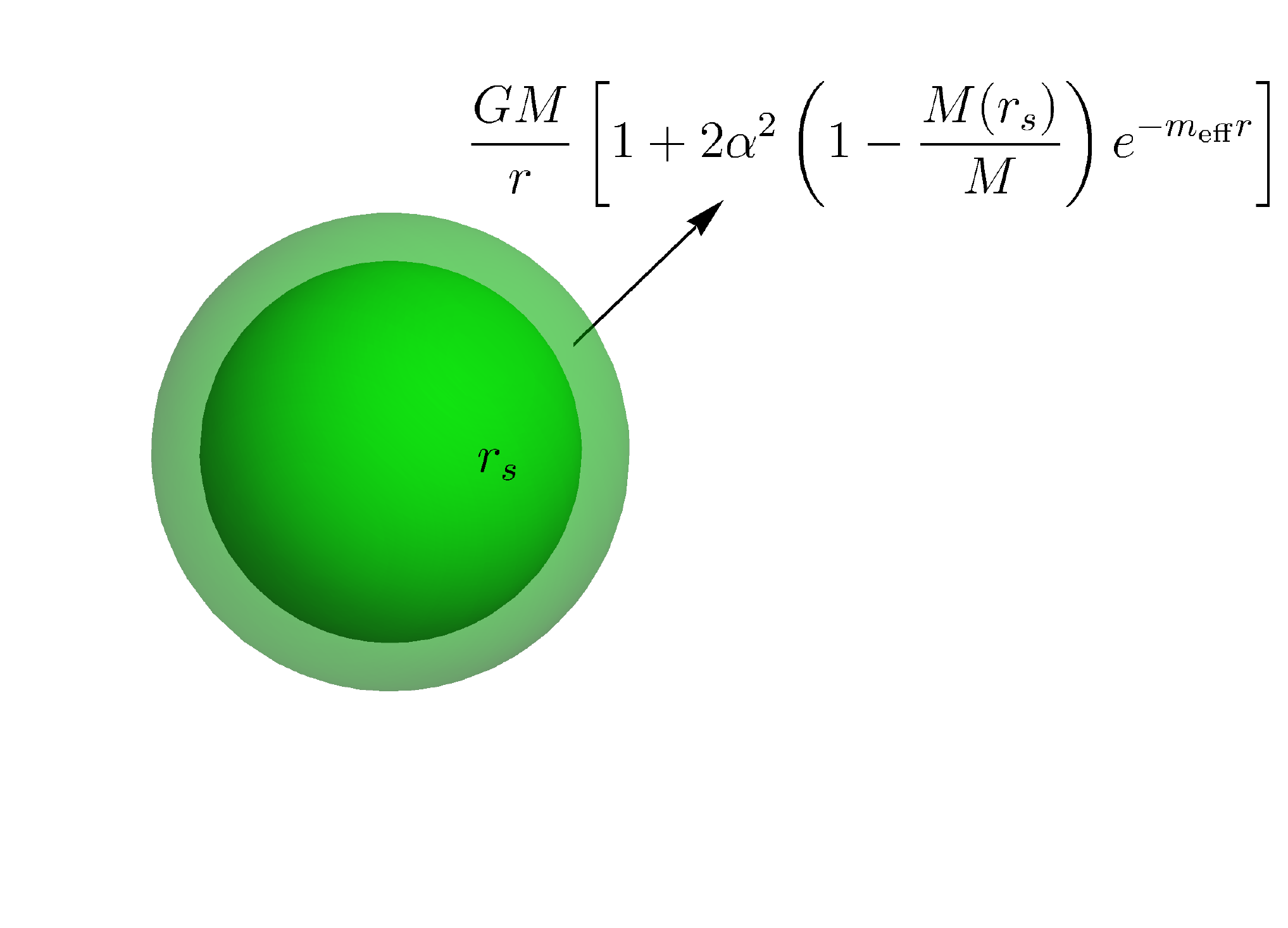}}
\caption{Chameleon screening. Only the mass inside the screening radius $\rs$ contributes to the fifth-force.}\label{fig:cham_screen}
\end{figure}

Astrophysically, the chameleon profile of a spherically-symmetric object of mass $M$ and radius $R$ is not sourced by the object's mass but rather by the mass inside a shell near the surface, a phenomenon that has been dubbed \emph{the thin-shell effect}. This is depicted in figure \ref{fig:cham_screen}. The reason for this is the following: deep inside the object, the field minimizes its effective potential corresponding to the ambient density but, as one moves away from the center, the field must eventually roll in order to begin to asymptote towards the minimum at the density of the medium in which the object is immersed (galactic densities or cosmological densities depending on the situation). The field can only roll once the density is low enough so that its effective mass is light enough. The radius at which this happens is typically called the \emph{screening radius} $r_{\rm s}$, and only the mass inside the screening radius sources a modification of the Newtonian potential, which is given by
\begin{equation}\label{eq:Vcham}
V(r)=\frac{GM}{r}\left[1+2\alpha^2\left(1-\frac{M(r_{\rm s})}{M}\right)e^{-m\eff r}\right].
\end{equation}
Objects for which $r_{\rm s}\approx R$ have drastically suppressed Yuakawa forces since $M(r_{\rm s})\approx M$ whereas those where $r_{\rm s}\ll R$ have strong enhancements. These two situations are referred to as \emph{screened} and \emph{unscreened} respectively. The screening radius of an object can be determined from the relation \cite{Sakstein:2013pda,Sakstein:2015oqa,Zhang:2016njn,Burrage:2017qrf}
\begin{equation}\label{eq:screenradcham}
\frac{\phi_{\rm min}^{\rm BG}}{2\alpha\mpl}=4\pi G \int_{\rs}^R r'\rho(r')\dd r',
\end{equation}
where $\phi_{\rm min}^{\rm BG}$ is the asymptotic (background) value of the field far from the object. If equation \eqref{eq:screenradcham} has no solutions then $\rs=0$ and the object is fully unscreened. 

Chameleon theories violate the weak equivalence principle \cite{Hui:2009kc}. Indeed, one can define a scalar charge for an object 
\begin{equation}\label{eq:chargedef}
Q_i= M_i\left(1-\frac{M_i(r^i_{\rm s})}{M_i}\right)
\end{equation}
so that the force on an object due to an externally applied chameleon field is $F_{\rm cham}=\alpha Q_i\nabla\phi^{\rm ext}$ (this is analogous to the `gravitational charge' $M$ so that $F_{\rm grav}=M\nabla\pn^{\rm ext}$ where $\pn^{\rm ext}$ is an externally applied Newtonian potential). {Two objects of different masses and internal compositions will have different scalar charges and will therefore fall at different rates in an externally applied chameleon field, signifying a breakdown of the weak equivalence principle (WEP). The chameleon force between two bodies, $A$ and $B$, is \cite{Mota:2006ed,Mota:2006fz}
\begin{equation}\label{eq:2force}
F_{AB}=-\frac{GM_AM_B}{r^2}\left(1+2\alpha^2 Q_A Q_B e^{-m\eff r}\right)
\end{equation}
and as a result of this the PPN parameter $\gamma$ is (see Appendix \ref{sec:appendix} for the derivation of this result and \cite{Hees:2011mu,Scharer:2014kya,Zhang:2016njn,Burrage:2017qrf} for other approaches)
\begin{equation}\label{eq:PPNcham}
\gamma=2\left[1+2\alpha^2Q_AQ_Be^{-m\eff r}\right]^{-1} -1.
\end{equation}
In this formula, body $A$ is the body responsible for the deflection/time delay of light and body $B$ is a separate body used to measure the mass of body $A$. For example, for light bending by the Sun one would take $A$ as the Sun and $B$ as the Earth. See Appendix \ref{sec:appendix} for more details.
}

\subsubsection{Symmetron Screening}

\begin{figure}
{\includegraphics[width=0.4\textwidth]{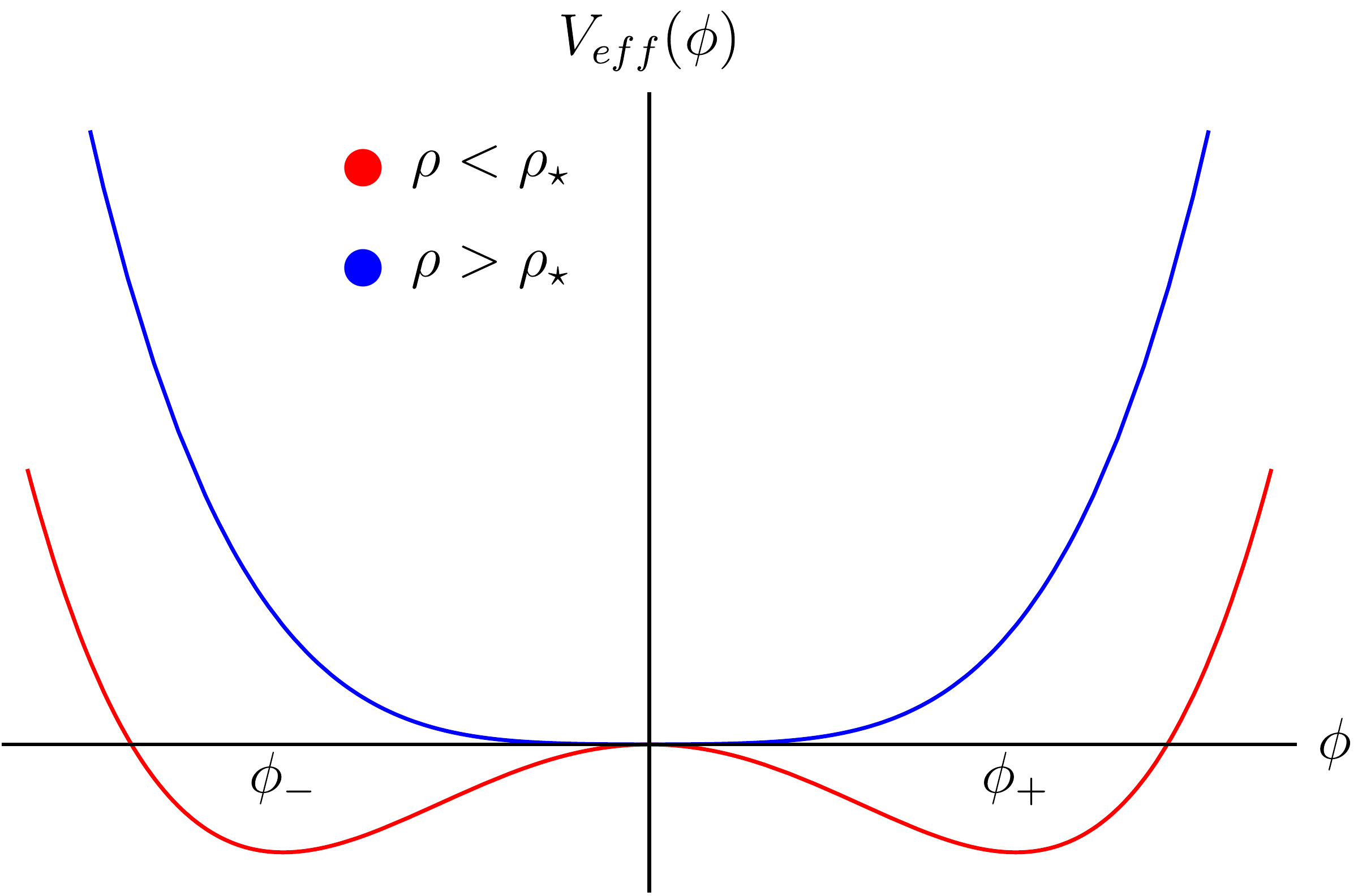}}
\caption{The symmetron effective potential. }\label{fig:symmveff}
\end{figure}

The symmetron model \cite{Hinterbichler:2010es} screens in a similar fashion to the chameleon---in the sense that it utilizes a mechanism similar to the thin-shell effect---but differs on how this is achieved. Instead of having a large mass inside the screening radius, the symmetron has a light mass (in all environments) but an environmentally-dependent coupling to matter that becomes zero. It's equation of motion is
\begin{equation}
\nabla^2\phi=\frac{\dd V_{\rm eff}}{\dd\phi}
\end{equation}
where the effective potential is
\begin{equation}\label{eq:symmveff}
V\eff(\phi)=-\frac{\mu^2}{2}\left(1-\frac{\rho}{\mu^2M_{\rm s}^2}\right)\phi^2+\frac{\lambda}{4}\phi^4,
\end{equation}
which is sketched in figure \ref{fig:symmveff}. This represents a field with a tachyonic mass $\mu$ and a field-dependent coupling to matter
\begin{equation}
\alpha(\phi)=\frac{\mpl \phi}{M_{\rm s}^2}.
\end{equation}
The effective potential can have two shapes depending on the magnitude of the ambient density: either there is a single minimum at $\phi=0$ when 
\begin{equation}
\label{eq:rhostar}
\rho>\rho_\star\equiv \mu^2 M_{\rm s}^2
\end{equation}
or there are two degenerate minima at
\begin{equation}
\phi=\phi_\pm\approx\pm\frac{\mu}{\sqrt{\lambda}}
\end{equation}
when $\rho<\rho_\star$. In both cases, the mass about this minimum is $m\eff^2=V\eff''(\phi)\sim\mu^2$ so that the mass does not vary significantly with density. When $\rho>\rho_\star$ the coupling vanishes identically since $\phi=0$ whereas when $\rho<\rho_\star$ the coupling is
\begin{equation}\label{eq:alpha0}
\alpha_0\equiv|\alpha(\phi_\pm)|=\frac{\mu\mpl}{\sqrt{\lambda} M^2}.
\end{equation}
The screening then works as follows: Given a spherical object embedded in a larger background of lower density ($\rho<\rho_\star$), the field will lie at $\phi=0$ at the center (provided the density $\rho>\rho_\star$ at some radius) and will remain here until the screening radius, at which point it begins to asymptote to $\phi_\pm$. When $r<\rs$ the coupling is zero and there is no fifth-force but when $r>\rs$ the coupling is $\alpha_0$ (given in \eqref{eq:alpha0}) and one finds, outside the object, 
\begin{equation}\label{eq:Vsym}
V(r)=\frac{GM}{r}\left[1+2\alpha_0^2\left(1-\frac{M(r_{\rm s})}{M}\right)e^{-\mu r}\right].
\end{equation}

{Like Chameleons, symmetrons also violate the WEP and one has precisely the same scalar charge as defined in equation \eqref{eq:chargedef} so that the force between two bodies is given by equation \eqref{eq:2force} with $m\eff=\mu$ and $\alpha\rightarrow\alpha_0$.} The PPN parameter $\gamma$ is, similarly, 
\begin{equation}\label{eq:PPNsym}
\gamma=2\left[1+2\alpha_0^2Q_AQ_Be^{-\mu r}\right]^{-1} -1.\end{equation}
where one again takes $r$ to be the typical length-scale of the experiment. The screening radius can be found by evaluating \cite{Burrage:2017qrf}
\begin{equation}\label{eq:screenradsymm}
M_{\rm s}^2= \int_{\rs}^R r'\rho(r')\dd r'.
\end{equation}
If there is no solution then $\rs=0$ and the object is fully unscreened. For the Sun, this is the case when $M_{\rm s}\gsim2.8\times10^{16}$ GeV and for the Earth one finds this is the case when $M_{\rm s}\gsim 6.68\times10^{15}$ GeV. 

\subsubsection{Vainshtein Screening: Galileons}

The Vainshtein mechanism \cite{Vainshtein:1972sx} is very general and is found ubiquitously in theories of massive gravity \cite{deRham:2010kj}, braneworld models \cite{Dvali:2000hr}, and general scalar-tensor theories \cite{Kimura:2011dc,Koyama:2013paa,Kobayashi:2014ida,Koyama:2015oma,Sakstein:2015zoa,Sakstein:2015aac,Sakstein:2015aqx,Sakstein:2016ggl}. In this paper, we will illustrate it by considering two simple and well-studied models that have become quintessential paradigms for the Vainshtein mechanism, the cubic galileon\footnote{{Note that we have taken the scalar to be dimensionless in contrast to the chameleon and symmetron scalars in order to match the conventions of reference \cite{Sakstein:2017bws}.}} \cite{Nicolis:2008in}
\begin{equation}\label{eq:cubicgal}
\nabla^2\phi + \frac{r_c^2}{3}\left[(\nabla^2\phi)^2-\nabla_i\nabla_j\phi\nabla^i\nabla^j\phi\right] =8\pi\alpha G\rho,
\end{equation}
and the quartic galileon
\begin{align}
\nabla^2\phi& + \frac{r_c^4}{4}\left[(\nabla^2\phi)^3-\nabla^2\phi\nabla_i\nabla_j\phi\nabla^i\nabla^j\phi\right.\nonumber\\&\left.+2\nabla_i\nabla_j\phi\nabla^j\nabla^k\phi\nabla_k\nabla^i\phi\right] =8\pi\alpha G\rho.\label{eq:quarticgal}
\end{align}
The new parameter $r_c$ is referred to as the \emph{crossover scale}\footnote{The cubic galileon is a certain limit of five-dimensional brane world models and this scale determines when the extra dimension is important. In the case of the quartic galileon there is no analog but we use the same symbol for the new parameter for the sake of consistency.}.
In each case, one can see that the left hand side of the equation of motion contains a Poisson term and a non-linear term. The relative importance of each term is determined by the \emph{Vainshtein radius}
\begin{align}\label{eq:rv}
r_{\rm V}^3=
  \begin{cases}
            \frac{4}{3} \alpha GMr_c^2,  &\quad \textrm{cubic galileon}\\
   \sqrt{2}\alpha GMr_c^2, & \quad \textrm{quartic galileon}
  \end{cases}.
\end{align}
When $r\gg r_{\rm V}$ the Poisson term dominates so that the field is Brans-Dicke-like and one has $\oo(\alpha^2)$ fifth-forces. On the other hand, when $r\ll r_{\rm V}$ the non-linear terms are dominant and one finds a total force
\begin{align}\label{eq:galileonforce}
F=\frac{GM}{r^2}\left[1+2\alpha^2\left(\frac{r}{\rv}\right)^p\right],
\end{align}
where $p=3/2$ for the cubic galileon and $p=2$ for the quartic. One can see that deviations from the inverse-square law are highly suppressed by powers of $r/\rv$ for distances inside the Vainshtein radius. For the solar system, the relevant Vainshtein radius is that of the Sun, which is of order 100 pc, and so deviations from GR are highly-screened in the solar system. Cosmologically, galileons can self-accelerate without the need for a cosmological constant provided that $r_c\sim 6000$ Mpc \cite{Schmidt:2010jr}. Galileons with smaller values of $r_c$ are not dominant cosmologically but represent new and interesting potential modifications of gravity that have yet to be well-constrained. Unlike chameleons and symmetrons, galileons obey the weak equivalence principle i.e. $Q=M$ \cite{Hui:2009kc}, although non-linear effects mean that this may not be the case for two or more extended bodies in close proximity \cite{Hiramatsu:2012xj}. 

\section{Present and Future Space-Based Experiments}\label{sec:missions}

In this section we briefly describe the experiments we will use to constrain the screened modified gravity models presented above. In particular, we will indicate the relevant tests for these theories and state the predicted precision with which the appropriate parameters can be measured.

\subsection{Cassini}

Primarily a mission to study the physics of Saturn, the Cassini satellite and Earth were in conjunction in 2002, allowing for a test of GR using the Shapiro time delay effect. Signals sent from Earth to the satellite (en route to it's future host planet at the time) that passed the Sun with different impact parameters were able to overcome the noise due to the solar corona and allow an accurate measurement of the time delay. The resulting constraint on the PPN parameter $\gamma$, $|\gamma-1|<2.1\times10^{-5}$, is currently the strongest bound on this parameter to date.

\subsection{Lunar Laser Ranging}

Laser ranging to five reflectors placed on the moon during the Apollo and Lunokhod missions can measure the relative distance between the Earth and the Moon with mm precision. This is achieved by measuring the round-trip time for short laser pulses aimed at these reflectors, a technique known as lunar laser ranging (LLR). The incredibly high precision has allowed for tests of general relativity at the Earth-Moon distance ($10^{10}$ cm). In particular, the time-variation of Newton's constant has been measured to $\dot{G}/G<6\times10^{-13}$ yr$^{-1}$, the inverse-square law has been verified with a precision $\delta V/V<2.1\times10^{-11}$, and the equivalence principle has been tested to $10^{-13}$ \cite{Williams:2004qba,Williams:2005rv,Turyshev:2006gm}. In the latter case, the bound refers to the differential acceleration between the Earth and Moon towards the Sun $(a_\oplus-a_{\leftmoon})/a_{\rm N}$, where $a_{\rm N}$ is the Newtonian acceleration.

\subsection{Phobos Laser Ranging}

Building on the success of LLR, it has been proposed to land a pulsed laser transponder on the surface of Phobos \cite{Turyshev:2010gk}. The resulting Phobos laser ranging (PLR) program would be able to achieve mm-level accuracies at the Earth-Mars distance (1.5 AU). Earth and Mars would be in conjunction after 1.5 years, with a second and third conjunction in three and six years respectively. {At conjunction, the laser pulses would pass close to the Sun and experience a strong Shapiro time delay effect due to the warping of space-time.} This would allow for constraints on the PPN parameter $\gamma$ to $10^{-7}$--$10^{-8}$ levels (the latter could be achieved after three conjunctions). Additionally, the time-variation of $G$ could be measured to $\dot{G}/G\lsim 10^{-15}$ yr$^{-1}$ and the inverse-square law could be tested to $10^{-11}$ at the Earth-Mars distance. Not only does this allow for tests of general relativity on small scales but theories such as galileons that predict forces that increase with distance could be constrained to new levels. The equivalence principle could be tested to the $10^{-15}$ level using the Earth-Mars-Sun-Jupiter system \cite{Anderson:1995df}.

\subsection{LATOR}

The Laser Astrometric Test of Relativity (LATOR) \cite{Turyshev:2004ky,Turyshev:2005ux,Turyshev:2005aw,Plowman:2005fb,Turyshev:2006wg,Turyshev:2005jf,Turyshev:2007pt} aims to place two microsatellites in heliocentric orbits on the far side of the Sun with orbital radius 1 AU. Lasers placed on the satellites will send light signals to an optical interferometer placed on the international space station (ISS). The line of sight of each microsatellite passes at close but different distances to the Sun so that the entire configuration forms a triangle. If there were no warping of space-time by the Sun then the geometry of this triangle would be exactly Euclidean but the warping leads to departures from this, which manifests as a deflection of the laser signal. Measuring the amount by which the properties of the triangle deviate from their Euclidean values therefore probes the PPN parameter $\gamma$, which will be measured to an accuracy $|\gamma-1|\sim2.7\times10^{-9}$. 

\begin{figure*}\centering
{\includegraphics[width=0.44\textwidth,valign=t]{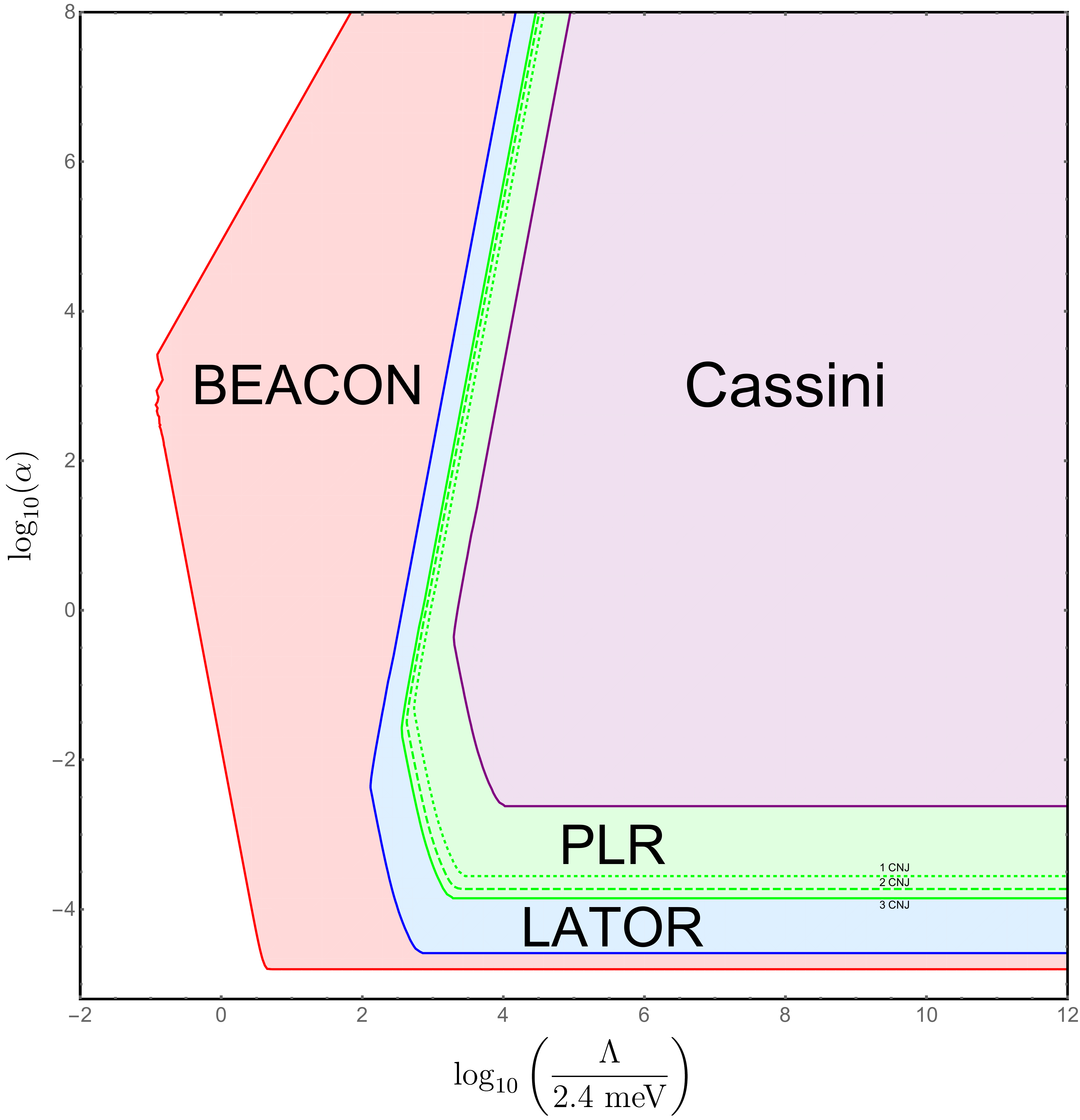}}
{\includegraphics[width=0.45\textwidth,valign=t]{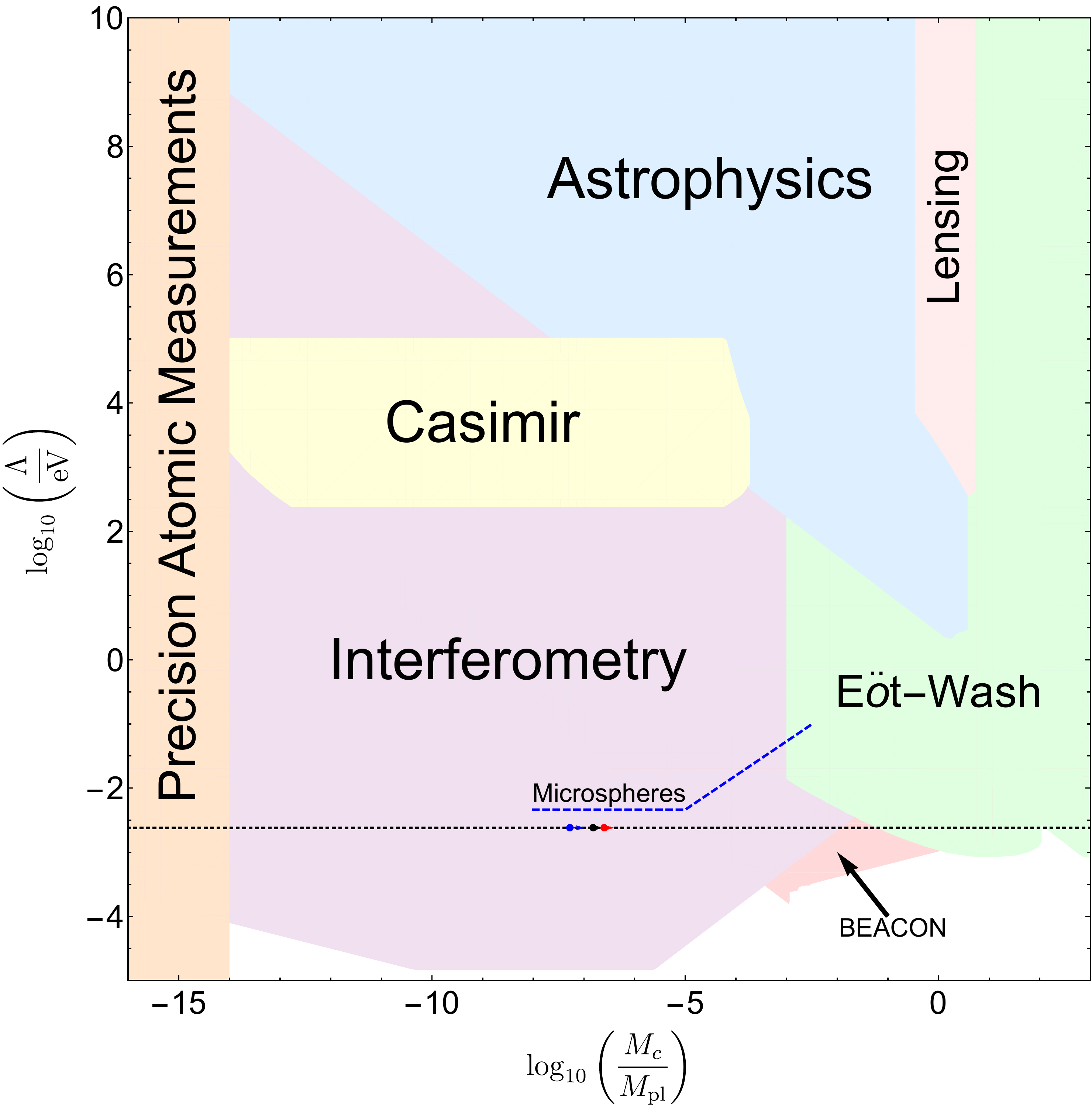}}
\caption{The potential exclusion range for $n=1$ chameleon models. \emph{Left Panel}: The region excluded in the $\Lambda$--$\alpha$ plane by future space based tests of the PPN parameter $\gamma$; the exclusion range due to each experiment is indicated in the figure. In the case of PLR, the green region indicates the constraints that could be obtained if three conjunctions are achieved; the corresponding region for one and two conjunctions are shown using the dotted and dashed green lines respectively. Note that we have normalized the chameleon mass-scale $\Lambda$ to the dark energy scale so that a value of zero indicates that the chameleon and dark energy may have a common origin. \emph{Right Panel}: Comparison with other chameleon bounds coming from the experiments labelled in the figure. In this case we have normalized $\Lambda$ in units of eV and have translated the bounds into $M_c=\mpl/\alpha$ in order to conform with conventions in the experimental literature. The black dotted line shows the dark energy scale and the colored arrows indicate lower bounds on $M_c$ coming from neutron bouncing and interferometry experiments. See \cite{Burrage:2017qrf} for a description of each of these experiments, and the resulting bounds.}\label{fig:gamma}
\end{figure*}

\subsection{GTDM}

The gravitational time delay measurement (GTDM) experiment \cite{Ashby2008} proposes to measure the Shapiro time delay effect using a similar configuration to LATOR, the difference being that laser ranging between two drag-free satellites, one at the L1 Lagrange point of the Earth-Sun system and one in a LATOR-type orbit, is to be used. This will measure the PPN parameter $\gamma$ to $|\gamma-1|<2\times10^{-8}$. We will not consider GTDM here since this will not be as strong as the LATOR constraint, which operates at the same distances. 

\subsection{BEACON}

Rather than measuring the space-time curvature sourced by the Sun, the Beyond Einstein Advanced Coherent Optical Network (BEACON) \cite{Turyshev:2008rh} will attempt to measure the space-time warping by the Earth. Four small satellites will be placed in circular orbits around the Earth (at distances of order $8\times10^{4}$ km) in a trapezoidal configuration. Laser transceivers on each satellite will allow the distances between any two satellites to be measured with high precision. The laser beams of opposite satellites passes close to the Earth and therefore the signal picks up a time delay due to the warping of space-time. Modulating the position of one spacecraft relative to the others changes the impact parameter and will therefore allow a measurement of the PPN parameter $\gamma$, which will be measured to an accuracy $|\gamma-1|=10^{-9}$.

\section{Potential Constraints}
\label{sec:constraints}
\subsection{Chameleons and Symmetrons}

Since Chameleons and Symmetrons screen in a qualitatively similar manner, we will consider potential tests of both simultaneously. Whereas chameleons and symmetrons do give rise to deviations in the inverse-square law, this is only the case for a narrow range of parameters (where the effective mass is of order the Earth-Moon or Earth-Mars distance for LLR and PLR respectively). Furthermore, it is likely that the field sourced by Mars will environmentally-screen Phobos (see \cite{Hui:2009kc} for a discussion on this) and a modeling of this effect is very complicated due to the high degree of non-linearity in the system \cite{Mota:2006ed,Mota:2006fz}. For this reason, tests using the PPN parameter $\gamma$ are cleaner and cover a larger range of parameter space; we will therefore focus on the bounds that future tests of this parameter place on chameleon and symmetron theories.

The tests described above will constrain $\gamma$ using the space-time warping due to either the Sun (CASSINI, PLR, LATOR) or the Earth (BEACON) {using either the Shapiro time delay effect or by measuring the deflection of light. The PPN parameter $\gamma$ is given in equations \eqref{eq:PPNcham} and \eqref{eq:PPNsym} for chameleon and symmetron models respectively. } {In the case of Cassini, PLR, and LATOR, body $A$ is the Sun and body $B$ is the Earth since its orbital dynamics are used to measure the Sun's mass. For BEACON, body $A$ is the Earth and body $B$ is the LAGEOS satellite, which has been used to make a measurement of the geocentric gravitational constant \cite{GRL:GRL909}. In practice, the LAGEOS satellite is fully unscreened (its Newtonian potential $GM/Rc^2$ is of order $10^{-25}$) and hence acts like a point particle. For this reason $Q\approx1$ for the LAGEOS satellite and BEACON is a better probe than the other tests considered here since there is only one factor of the scalar charge. } 

{We find the screening radius for the Sun, Earth, and LAGEOS satellite by integrating equation \eqref{eq:screenradcham} (\eqref{eq:screenradsymm} for symmetrons) given a relevant density profile. We then use these in equation \eqref{eq:PPNcham} (\eqref{eq:PPNsym} for symmetrons) to calculate the range of parameters for which the predicted value of $\gamma$ will exceed the projected bound coming from each experiment. In the case of the Sun, we use the solar density profile of \cite{Bahcall:2004pz}.  {We take the dark matter density in the solar neighborhood to be $\rho_{\rm DM}=0.324$ GeV/cm$^3$ ($6\times10^{-25}$g/cm$^3$) \cite{Read:2014qva}, which sets the background value of the field $\phi^{\rm BG}_{\rm min}$ {for chameleons} given a set of parameters}. {For the Earth, we assume a mean density of $5.51$ g/cm$^3$ and for the LAGEOS satellite we calculate the mean density $\rho=3M/(4\pi R^3)\approx 0.45$ g/cm$^3$ (assumed constant) using the mass ($407$ kg) and radius ($60$ cm).}}

{In the left panel of figure \ref{fig:gamma} we show the potential regions of parameter space that could be excluded for $n=1$ chameleon models and compare these with current experimental constraints taken from \cite{Burrage:2017qrf}. Note that it is common in the experimental literature to write $\alpha=\mpl/M_c$ and so we do the same here for comparative purposes. One can see that only BEACON will be competitive with current experimental searches. It is interesting that the region constrained by BEACON is the region where $\Lambda$ is of order the dark energy scale. BEACON therefore has the possibility to rule out $n=1$ chameleon models that may have a common origin with dark energy. }

\begin{figure}
{\includegraphics[width=0.45\textwidth]{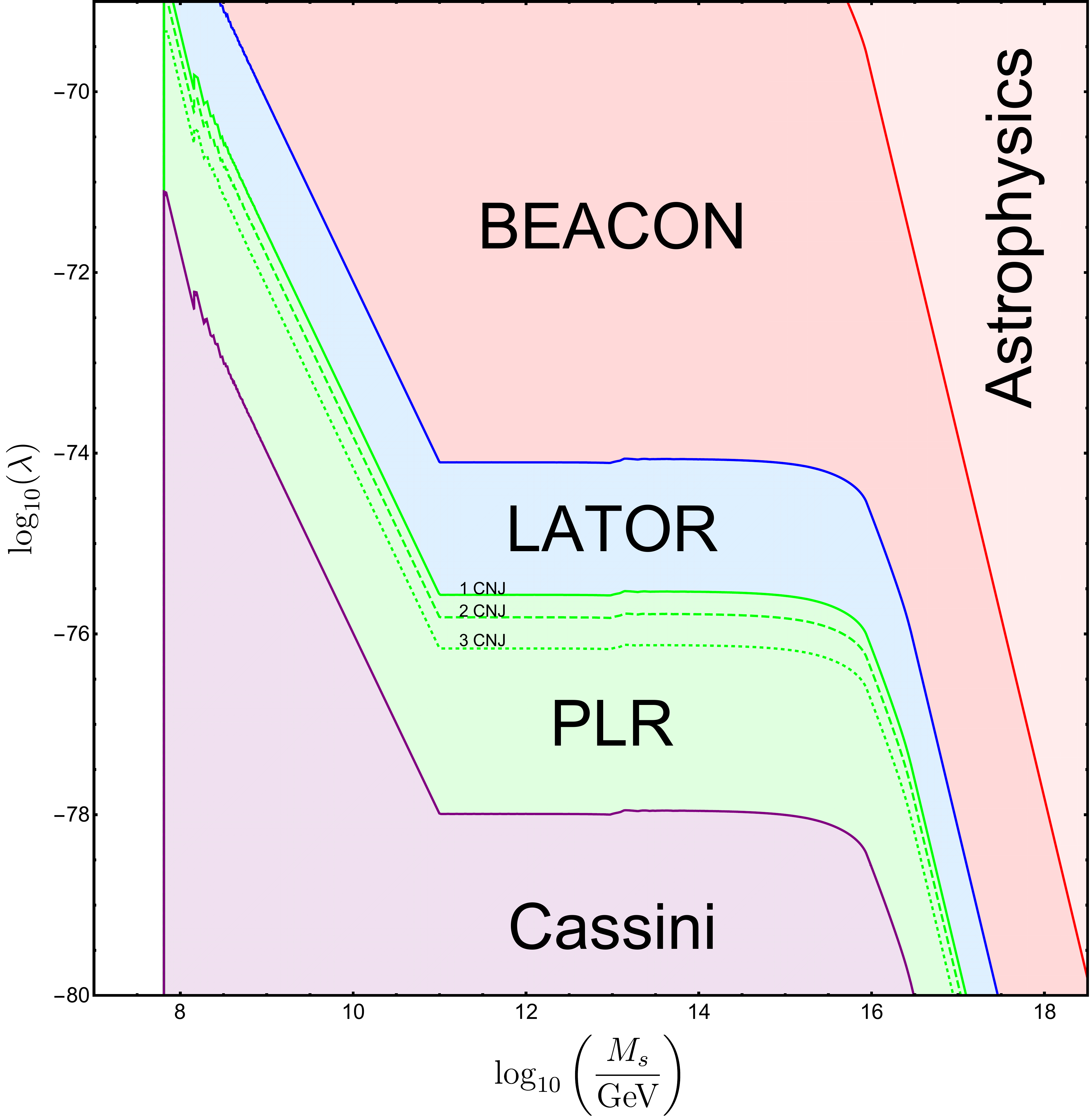}}
\caption{Current and potential constraints on the symmetron model with $\mu=8\times10^{-18}$ eV from space based tests of the PPN parameter $\gamma$. The shaded regions correspond to the current or future experiment indicated in the figure. The bounds from astrophysical searches are also shown.  }\label{fig:symcons}
\end{figure}

{In figure \ref{fig:symcons} we show the region of symmetron parameter space that could potentially be excluded. We focus on models with $\mu=8\times10^{-18}$ eV so that the force range is larger than one AU.} Note that the parameter range in is very different those considered in laboratory tests \cite{Upadhye:2012rc,Brax:2016wjk,Burrage:2016rkv,Elder:2016yxm,Jaffe:2016fsh}. Symmetrons are far less constrained than chameleons and there are typically large gaps in the parameter space separating constraints from laboratory and astrophysical tests \cite{Burrage:2017qrf}. In order to be able to probe screened fifth-forces in a laboratory setting, the Compton wavelength must be of order (or smaller than) the width of the walls of the vacuum chamber in which the experiment is performed. Chameleons can vary their mass over many orders of magnitude and therefore different laboratory tests can probe a complementary range of parameter space whereas symmetrons have a fixed mass of $\mathcal{O}(\mu)$ and hence the range of parameters that can be probed is limited. The parameters we consider in figure \ref{fig:gamma} are adapted to the solar system rather than laboratory tests.

One can see from figure \ref{fig:symcons} that the same region that is constrained by solar system tests is also probed by astrophysical tests using distance indicators and rotation curves \cite{Davis:2011qf,Jain:2012tn,Sakstein:2013pda,Sakstein:2014nfa,Vikram:2014uza,Sakstein:2015oqa}. In this case, solar system tests constrain a complementary region of parameter space to astrophysical tests, and therefore future space-based tests will cover a currently unconstrained region. Note that it is not possible to extend the astrophysical bounds to arbitrarily small values of $M$ because these tests require dwarf galaxies in cosmic voids to be unscreened. According to equation \ref{eq:screenradsymm}, these galaxies would become screened at small $M$ and, in fact, the minimum value of $M$ constrained in the figure is close to the threshold for the onset of screening in dwarf galaxies.

\subsection{Cubic and Quartic Galileons}

\begin{figure*}\centering
\stackunder{
{\includegraphics[width=0.45\textwidth]{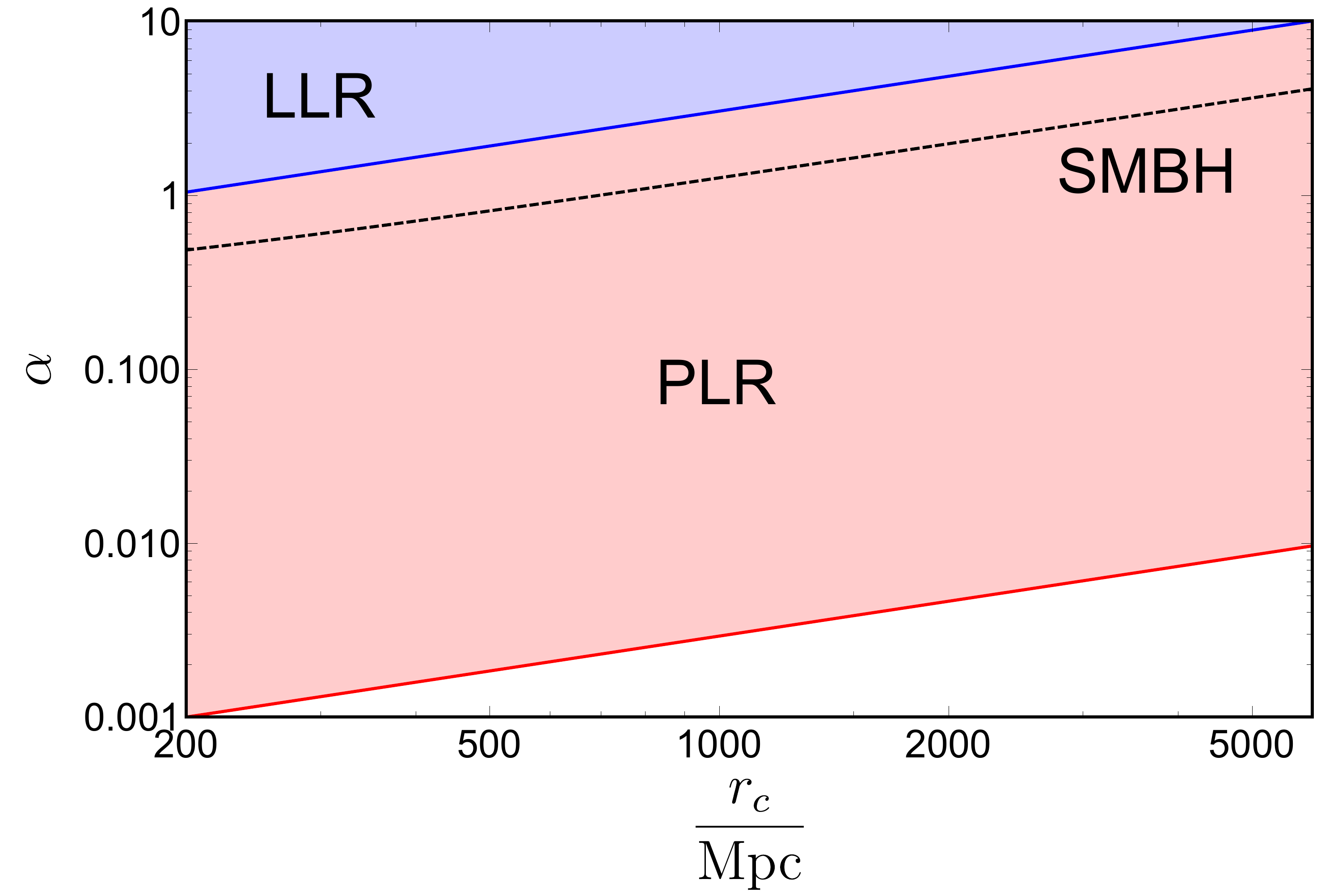}}}{$\phantom{xxxxxx}$Cubic Galileon}
\stackunder{
{\includegraphics[width=0.45\textwidth]{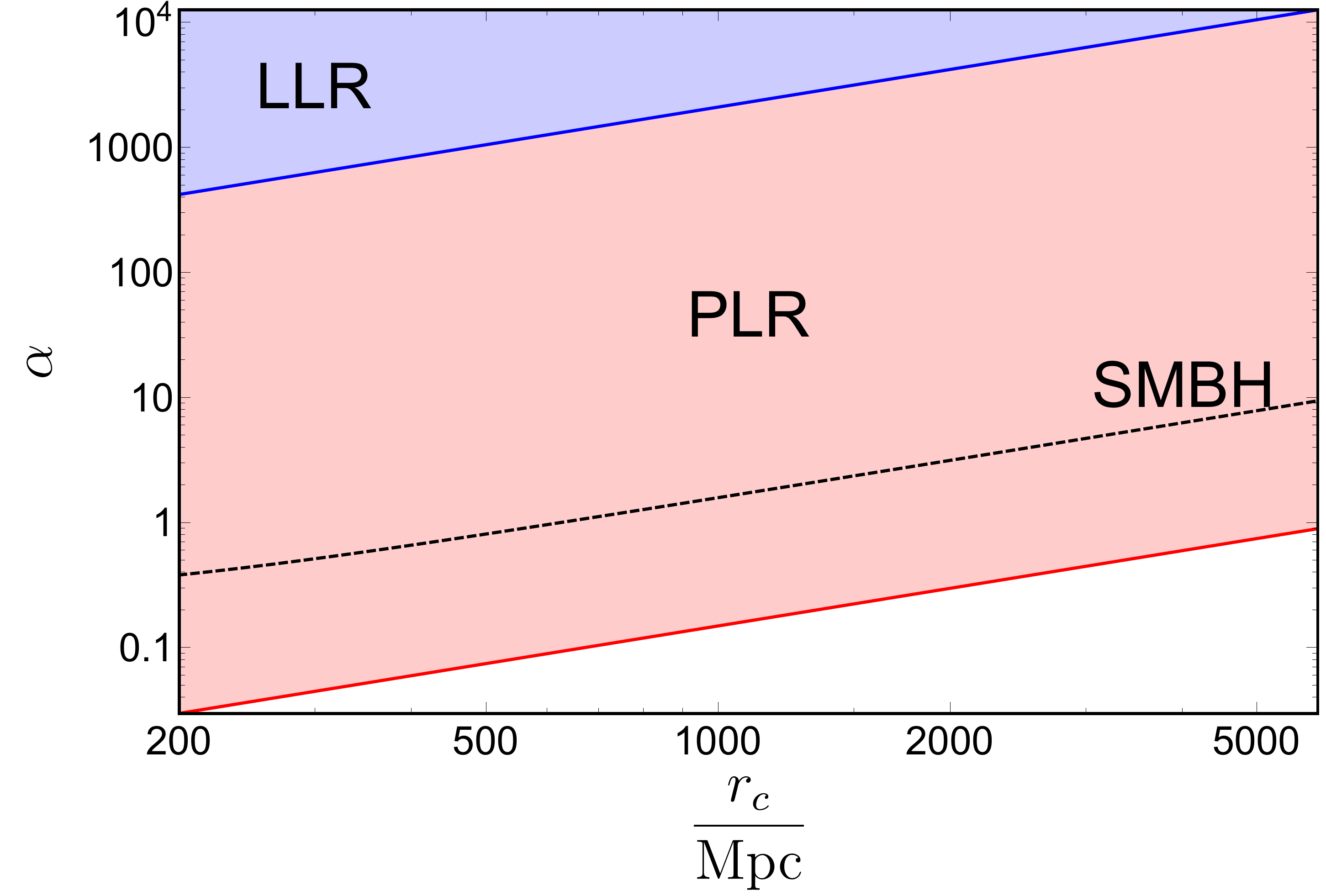}}}{$\phantom{xxxxx}$Quartic Galileon}
\caption{Current and potential constraints on the cubic (left panel) and quartic (right panel) galileon models. The blue regions are the current bounds from LLR and the region above the black dashed line is excluded by the SMBH tests of reference \cite{Sakstein:2017bws}. The red region shows the parameter space that could be potentially ruled out with PLR.  }\label{fig:LRR_galileon}
\end{figure*}

Galileon theories produce deviations in the inverse-square law (see equation \eqref{eq:galileonforce}) which can be tested with laser ranging. Since the galileon force generated by the Earth scales with distance to some positive power, it is stronger at the Earth-Mars distance than the Earth-Moon distance. PLR will therefore improve the bounds over the current LLR bounds \cite{Dvali:2002vf,Khoury:2013tda}. According to equation \eqref{eq:galileonforce}, one has
\begin{equation}
\frac{\delta V}{V}=\left(\frac{r}{\rv}\right)^p,
\end{equation}
with $\rv$ given in equation \eqref{eq:rv} and where $p=3/2,$ $2$ for the cubic and quartic galileon respectively. {$r$ should be taken to be the Earth-Mars distance in the case of PLR.} Demanding that $\delta V/V$ is less than the bound reported from LLR and the predicted sensitivity of PLR (both $10^{-11}$ at the Earth-Moon and Earth-Mars distance respectively) we obtain the bounds (LLR) and predicted improvements (PLR) shown in figure \ref{fig:LRR_galileon}.

There have been few tests of galileon theories on small scales to date, due mainly to the efficiency of the Vainshtein mechanism. LLR yielded the strongest constraints until they were overtaken recently by tests using supermassive black holes (SMBHs) \cite{Sakstein:2017bws}. Galileons predict violations of the strong equivalence principle (SEP) so that black holes do not couple to external galileon fields whereas non-relativistic matter does \cite{Hui:2012qt,Hui:2012jb}. The acceleration of a galaxy infalling into a massive cluster receives a large but subdominant contribution from the galileon field of the cluster\footnote{This is because extended distributions do not screen as efficiently as point sources in galileon theories \cite{Sakstein:2017bws}.} that the SMBH at its center does not feel. For this reason, as the galaxy falls towards the center of the cluster the black hole begins to lag behind and is eventually stabilized by the restoring force of the baryons left at the galactic center. This results in an observable offset that can be as large as $\oo(\textrm{kpc})$. Reference \cite{Sakstein:2017bws} used the lack of an offset in the central SMBH of M87 (located in the Virgo cluster) to place the constraints that we also show in figure \ref{fig:LRR_galileon}. One can see that these are stronger than the current LLR bounds but that the bounds from PLR would supersede these. The bounds from PLR would therefore be the strongest bounds on galileons gravity models on small scales.

\subsection{Other Tests of Gravity in Space}

We end by discussing other possibilities for testing screened modified gravity in space that we will not forecast here due to technical difficulties with the theoretical modeling. Our goal is to point out that future effort in this direction could yield fruitful results. 

\subsubsection{Tests of the Time-Variation of Newton's Constant}

All three of the screening mechanisms mentioned here predict that Newton's constant is time-dependent. This is primarily because the asymptotic value of the field is given by the cosmological or galactic field value (depending on the parameters one chooses). For chameleon and symmetron models, the galaxy (and therefore the Sun) must be unscreened for the former case to be relevant and, in the latter case, the time-dependence of the field is very model-dependent and requires detailed N-body simulations or excursion set methods to predict \cite{Li:2011qda}. For this reason, it is difficult to use the time-variation of $G$ to constrain symmetron and chameleon models.

Galileon models predict a strong variation in the time-dependence of $G$ \cite{Li:2013tda,Babichev:2011iz}. Again, this time-dependence comes from the cosmological boundary conditions that make the coupling to matter $\alpha$ time-dependent. This time-dependence is fixed by the cosmology of the galileon and therefore constrains a combination of the fundamental model parameters, as well as $\dot{\phi}$, $H$, and $\Omega\mmm$. Since we are not interested in these parameters here we will not attempt to forecast how the improved bounds on $\dot{G}/G$ will impact galileon cosmology, but, for completeness, we will make a few pertinent observations. First, galileon models with a direct coupling to matter (that we study here) have not been well-studied in the context of cosmology; some references \cite{Chow:2009fm,Appleby:2011aa,Brax:2015dma} have studied the theoretical cosmology but no fiducial model has been proposed. Second, those that have been studied differ from the models studied in this work in that they produce self-acceleration by having a phantom quadratic kinetic term (i.e the term $\nabla^2\phi\rightarrow-\nabla^2\phi$ in the equations of motion \eqref{eq:cubicgal} and \eqref{eq:quarticgal}). These models are in tension with (but not excluded by) current cosmological data \cite{Renk:2017rzu} {and are strongly ruled out (except for very fine-tuned cases) by the recent LIGO/Virgo-Fermi observation of gravitational waves and a gamma ray burst from merging neutron stars, which constrains the difference in the speed of gravitons and photons to the $10^{-15}$ level \cite{Sakstein:2017xjx,Creminelli:2017sry,Ezquiaga:2017ekz,Baker:2017hug,Crisostomi:2017lbg,Langlois:2017dyl,Kreisch:2017uet}}. Finally, many different theories of gravity, including massive gravity \cite{deRham:2010kj,Hinterbichler:2011tt,deRham:2014zqa} and Horndeski theories (and their generalizations) \cite{Horndeski:1974wa,Deffayet:2009wt,Gleyzes:2014dya,Gleyzes:2014qga} reduce to the same galileon theories considered in this work on solar system scales but give very different cosmologies. For this reason, constraining the time-variation of $G$ does not constrain the fundamental parameters $\alpha$ and $r_c$ considered here. In many of these theories, the matching of small scales to a cosmological backgrounds presents a separate technical challenge because one needs to make sure to use a metric for the solar system that is consistent with cosmological asymptotics \cite{Babichev:2012re,Babichev:2016jom,Sakstein:2016lyj,Sakstein:2016oel}.  

\subsubsection{Tests of Other PPN Parameters}

In addition to $\gamma$, some of the tests mentioned above will also probe the PPN parameters $\beta$ and $\delta$ (for the first time in this case). The resulting bounds on these parameters will be weaker than for $\gamma$, and, since the relevant combination of parameters is the same for these additional parameters, no new information will be gained by constraining them\footnote{This is only true for screened modified gravity theories and is a result of the Newtonian scalar field ($\oo(v^2/c^2)$) being highly-suppressed. More general scalar-tensor theories will source post-Newtonian fields that are sensitive to higher-order corrections to $\alpha$, which will be additionally constrained by measuring $\beta$ and $\delta$. Additionally, Vainshtein screened theories have additional terms in the metric that are not captured by the PPN expansion alone \cite{Avilez-Lopez:2015dja} and it may be possible to constrain these. }. This conclusion may be different for theories that include a disformal coupling to matter \cite{Ip:2015qsa}.

\subsubsection{Tests of the Equivalence Principle}

\paragraph{The Strong Equivalence Principle:}

The Earth-Sun-Mars-Jupiter system allows for a novel test of the SEP \cite{Anderson:1995df} that could be performed using PLR \cite{Turyshev:2010gk}. Since chameleon and symmetron models violate the WEP, this system could be used to test these theories. In particular, all of these bodies will have different thin-shells (and hence scalar charges $Q$) and will therefore fall towards any given body at a different rate. The analysis by \cite{Anderson:1995df} involved integrating the equations of motion for the four-body system including higher-order effects such as tidal forces. Such an analysis would be more difficult for chameleon and symmetron models due to their non-linear nature. Any analytic solution would require different approximations because the regime of validity depends on the model parameters \cite{Mota:2006ed,Mota:2006fz}---for example, if the mass of the field becomes of order the separation between two bodies then superposition no longer holds---and deviations from spherical symmetry (including tidal effects) are harder (but not impossible \cite{Burrage:2014daa}) to model. In practice, it is likely that a numerical integration of the field equations will be necessary to find the resulting constraints.

Similarly, whereas the WEP is satisfied for a single extended object in galileon theories, two or more extended objects may violate the WEP due to the failure of superposition that results from the high degree of non-linearity in the equations of motion. It is likely that the theoretical modeling of this four-body system would be even harder for galileon theories. Their equations are harder to solve and may have multiple branches of solutions, deviations from spherical symmetry are poorly understood (except in other symmetric situations \cite{Bloomfield:2014zfa}), and it is not clear that perturbation theory works for these models \cite{Andrews:2013qva}. 

\paragraph{The Weak Equivalence Principle:}

There are several proposed experiments that will measure the WEP using accelerometers (of various design) orbiting the Earth \cite{Touboul:2012ui,Nobili:2012uj,Overduin:2012uk}. The perpetual free-fall of these accelerometers will allow for longer experiments. The satellite test of the equivalence principle (STEP) \cite{Overduin:2012uk} will reach a precision of $10^{-18}$, five orders of magnitude stronger than the current bounds from LLR ($10^{-13}$). In all cases, these experiments consist of a capsule in orbit around Earth with two test-bodies (typically cylinders that are designed to resemble spheres to high multipole moments) that free-fall towards the Earth, and accelerometers designed to measure any difference between the free-fall rates. It is unlikely that these experiments will constrain galileon theories because the capsule is a point mass to a good approximation but chameleons and symmetrons are very sensitive to the precise geometry of experimental chambers (see \cite{Burrage:2017qrf} for a review of experimental tests). The fact that these accelerometers operate with a highly non-symmetric geometry makes the theoretical modeling of the field profile very difficult and a numerical treatment would be necessary in order to make predictions. The paper that originally introduced chameleons estimated the range of parameters for which a STEP-like experiment would be unscreened \cite{Khoury:2003rn} by demanding that the capsule has no thin-shell, but going beyond this would require considerably more effort and so we do not attempt this here. It may be that a dedicated vacuum chamber in space with a geometry specifically tailored to optimize the chameleon and symmetron WEP violations would provide more stringent results than a detailed reanalysis of the current generation of planned and proposed accelerometers.

\section{Conclusions}
\label{sec:concs}

In this work we have explored the implications that current, planned and, proposed space-based tests of relativistic gravitation have for theories of gravity that include \emph{screening mechanisms}. Screening mechanisms use non-linear equations of motion to dynamically suppress deviations from GR in the solar system without the need to tune the theory parameters to negligible values. They therefore allow the theories to be relevant on cosmological scales, potentially allowing them to address the dark energy mystery. We have examined three well-studied and common paradigms for screening: chameleon, symmetron, and galileon models. (The latter models are paragons for Vainshtein screening, which occurs in a very broad class of scalar-tensor theories.) 

In the case of chameleon and symmetron models, which screen in a qualitatively similar manner using the thin-shell effect (see section \ref{sec:screening_mechanisms}), we have argued that space-based tests of the PPN parameter $\gamma$ using either laser ranging to Phobos (PLR) or optical networks (LATOR and BEACON) will provide the best constraints. {Only the potential BEACON bounds on chameleon models will probe into the region of parameter space not yet covered by current tests. BEACON has the ability to fill in the remaining region around the dark energy scale where chameleons and dark energy may have a common origin.} The bounds on symmetrons will be complementary to current bounds from astrophysical probes. For galileon models, the strongest constraints would come from testing the inverse-square law at interplanetary distances using PLR. In particular, tests of the inverse-square law would provide the strongest constraints on galileon models to date.

Finally, we have discussed whether or not the next generation of experiments aimed at testing the strong and weak equivalence principles in space could provide new and improved constraints. Ascertaining how strong these would be (if at all) is difficult due to uncertainties in the theoretical modeling of the both the four-body field profile and dynamics of the Earth-Sun-Jupiter-Mars system, and the proposed accelerometers that will be placed in orbit around Earth. In the former case, one simply needs to numerically solve the non-linear equations. In the latter, it is likely that only a small range of parameters can be probed, and it may be more fruitful to have a specifically designed vacuum chamber in space.

To paraphrase the paper that first introduced chameleons \cite{Khoury:2003aq}: 14 years later, screened modified gravity is still awaiting surprise tests for gravity in space.

\begin{acknowledgments}
{The author is indebted to the anonymous referee for their detailed reviewing of this manuscript, and to Justin Khoury for numerous useful discussions.} I would like to thank Bhuvnesh Jain, Mark Trodden, Alexander Vikman, and especially Slava Turyshev for comments on the manuscript. JS is supported by funds provided to the Center for Particle Cosmology by the University of Pennsylvania.
\end{acknowledgments}

\appendix

\section{Light Bending and Time Delay in Chameleon and Symmetron Theories}
\label{sec:appendix}

The purpose of this Appendix is to calculate the PPN parameter $\gamma$ that is relevant for chameleon and symmetron theories. We briefly review the pertinent theoretical aspects of these theories before deriving a value for $\gamma$. The reader is referred to \cite{Burrage:2017qrf} and references therein for further details. 

Chameleon and symmetron models are both scalar-tensor theories defined in the Einstein frame by 
\begin{equation}
S=\int\dd^4 x \sqrt{-g}\left[\frac{\mpl^2}{2}R(g)-\frac{1}{2}\nabla_\mu\phi\nabla^\mu\phi -V(\phi)\right]+S\mmm\left[\tg\right],
\end{equation}
where the Jordan frame metric $\tg_\nm$ is a Weyl rescaling of the Einstein frame metric $g_\nm$ by a conformal factor $A(\phi)$
\begin{equation}\label{eq:Weyl}
\tg_\nm=A^2(\phi)g_\nm.
\end{equation}
The coupling 
\begin{equation}
\alpha(\phi)=\mpl\frac{\dd\ln A(\phi)}{\dd\phi}
\end{equation}
and the specific model is set by the choice of $A(\phi)$ and $V(\phi)$. The specific model is unimportant for what follows.

Our starting point for the derivation is the PPN metric for a single body, which we will refer to as body A with mass $M_A$
\begin{align}\label{app:PPN_metric1}
\tilde{g}_{00}&=-1 +2 \frac{\GP M_A}{r}\\
\tg_{ij}&=\left(1+2\tilde{\gamma} \frac{\GP M_A}{r}\right)\delta_{ij},\label{app:PPN_metric2}
\end{align}
where we use tildes to refer to the Jordan frame metric, which governs the geodesics of point particles. The parameter that controls light bending/time delay measurements in this metric is $\tilde{\gamma}$. As we will see shortly, for theories that violate the WEP (such as ours) this is different from the parameter $\gamma$ constrained by measurements of these effects. We will refer to the quantity $\GP$ as the PPN gravitational constant because its value controls the size of effects computed using the PPN metric. It is distinct from the gravitational constant $G$ appearing in the action and the gravitational constant measured on Earth. The time delay and gravitational lensing of light is given by \cite{Will:2014kxa}
\begin{align}
\Delta t&=2\left(1+\tilde{\gamma}\right)\frac{\GP M_A}{c^3}F_1(b,x^\mu)\quad\textrm{and}\quad\label{eq:TD} \\
\delta \theta & = \left(\frac{1+\tilde{\gamma}}{2}\right)\frac{\GP M_A}{b c^2}F_2(b,x^\mu)\label{eq:LD}
\end{align}
where $b$ is the impact parameter and $F_1$ are geometric factors that depend on the geometry used to perform the measurement \cite{Shapiro:1964uw}. Their expressions are not necessary for what follows. 

We now wish to calculate the Jordan frame metric in PPN form. we first calculate the Einstein frame metric, defined by
\begin{align}
g_{00}&=-1+2\Phi\\
g_{ij}&=\left(1+2\Psi\right)\delta_{ij}.
\end{align}
One finds \cite{Hui:2009kc}
\begin{equation}
\nabla^2\Phi=\nabla^2\Psi=-4\pi G\rho_A
\end{equation}
so that
\begin{equation}
\Phi=\Psi=\frac{GM_A}{r}
\end{equation}
up to irrelevant integration constants set by the boundary conditions. For the scalar, we are interested in the regime where this body has some degree of screening. In this case, the equation inside the screening radius is
\begin{equation}
\nabla^2\phi=0
\end{equation}
while outside the screening radius it is 
\begin{equation}
\nabla^2\phi=8\pi\alpha G\rho_A\quad r>\rs,
\end{equation}
where $\alpha=\alpha(\phi_0)$, which is constant for chameleons and given by \eqref{eq:alpha0} for symmetrons. We have ignored the scalar's mass $m_{\eff}^2=V\eff''(\phi_0)$ ($\approx\mu^2$ for the symmetron) since we expect $m\eff R\ll1$ in the regime of interest but we can account for this by multiplying our final result by $e^{-m\eff (\phi_0)R}$ (more technical and cumbersome derivations find this factor \cite{Hees:2011mu}). The solution is then
\begin{equation}
\phi=\phi_0-2\alpha^2\frac{Q_AG M_A}{r},
\end{equation}
where the `scalar charge' of body A is
\begin{equation}\label{eq:appcharge}
Q_i=\left(1-\frac{M_A(\rs^A)}{M_A}\right).
\end{equation}

Transforming to the Jordan frame using equation \eqref{eq:Weyl} and expanding $A(\phi)$ to first order in $GM_A/r$ one finds equations \eqref{app:PPN_metric1} and \eqref{app:PPN_metric2} with
\begin{align}
\label{eq:GP} \GP&=G\left[1+2\alpha^2Q_A\right] \\
\tilde{\gamma} & = \frac{1-2\alpha^2Q_A}{1+ 2\alpha^2Q_A}. \label{eq:gam1}
\end{align}
Had we been dealing with a theory with no WEP violations our task would be complete since one could simply apply the constraints on $\gamma$ to \eqref{eq:gam1} but WEP violations imply that it is possible that $\GP$ can differ from the value of Newton's constant measured by local experiments. To see this, it is simpler to consider the product $GM$. In particular, consider measuring this combination using the orbital dynamics of a smaller body of mass $M_B$ orbiting the larger body sourcing this metric. This second body may have its own screening radius $\rs^B$ so that the force on this smaller body is
\begin{equation}
F=-\frac{GM}{r^2}\left[1+2\alpha^2Q_AQ_B\right]
\end{equation}
where we have once again ignored the mass of the scalar. The quantity that is measured in these theories is therefore
\begin{equation}
GM\left[1+2\alpha^2Q_AQ_B\right]\equiv (\GN M)_{\rm GR},
\end{equation}
 where $(\GN M)_{\rm GR}$ is the product of the mass and gravitational constant that one would infer in GR, or, rather, from Newtonian mechanics. It is this combination, in particular its numerical value, that must be used in equations \eqref{eq:TD} and \eqref{eq:LD} in order to correctly apply the constraints on $\gamma$. We therefore have
%\begin{widetext}
\begin{equation}
\gamma =  2\left[1+2\alpha^2Q_AQ_B\right]^{-1}-1.
\end{equation}
%\end{widetext}

\bibliography{ref}
\end{document}